\documentclass[%
 preprint,
 amsmath,amssymb,
 aps, physrev,
]{revtex4-2}
\usepackage[english,ngerman]{babel}
\usepackage{hyperref}
\usepackage[all]{hypcap}
\usepackage{subcaption}
\usepackage{mwe}
\usepackage{breqn}
\usepackage{amsmath}
\usepackage{float}
\usepackage[version=3]{mhchem}
\usepackage[section]{placeins}
\usepackage{graphicx}
\usepackage{dcolumn}
\usepackage{bm}
\usepackage[LGR,T1]{fontenc}
\usepackage{textgreek}

\DeclareUnicodeCharacter{2261}{=}
\renewcommand{\selectlanguage}[1]{} 
\begin{document}


\title{\textbf{Ab Initio studies on strain-dependant thermal transport properties of Graphene} 
}%
\author{Archishman Gupta}%
 \email{Contact author: archishman.gupta.cd.ece21@itbhu.ac.in}
 \affiliation{%
    Indian Institute of Technology (BHU) Varanasi.\\
    }%
\author{Ankit Arora}%
 \email{Contact author: ankit.ece@iitbhu.ac.in}
\affiliation{%
 Indian Institute of Technology (BHU) Varanasi.\\
}%

\date{\today}

\begin{abstract}
 In this work we present a comprehensive investigation of graphene’s thermal conductivity (\(\kappa\)) using first-principles density functional perturbation theory calculations, with a focus on the phonon and lattice vibrational properties underlying its superior heat transport capabilities. The study highlights the role of phonon frequencies, lifetimes and mode resolved contributions in determining graphene's thermal performance, emphasizing its high phonon group velocities and long mean free paths that contribute to thermal conductivity exceeding 3000 W/mK at room temperature.

The results are compared with other two-dimensional materials like silicene ($\kappa < 10$ W/mK) and MoS$_2$ ($\kappa \approx 83$ W/mK), to underline graphene's advantages in nanoscale applications. Here we report the concept of "velocity-lifetime trade off" and use it to explain graphene's excellent invariance to high tensile and compressive strains as it exhibits minimal variation in thermal conductivity, making it an ideal material for applications requiring stability in environments with strain variability and deformation. This study establishes graphene as a benchmark material for thermal transport in next-generation 2D channel FET devices and offers a roadmap for its optimization in practical applications.

\begin{description}
\item[keywords]
Phonons, thermal conductivity, transport, strain, vibrational analysis
 
\end{description}
\end{abstract}

\maketitle


\section{Introduction }

Thermal devices play a critical role in modern technology, enabling efficient heat management and energy conversion in applications ranging from microelectronics to thermoelectric systems. As the demand for miniaturized and high-performance devices continues to grow, effective thermal management becomes increasingly essential to ensure device stability, reliability, and efficiency. The key parameter governing the performance of thermal devices is Lattice thermal conductivity ($\kappa_l$), which dictates the material’s ability to transport heat. Achieving a balance between high thermal conductivity and low thermal diffusivity is crucial for optimizing heat dissipation and energy conversion.
\\

Conventional materials such as silicon, copper, and aluminium have been widely used in thermal devices due to their relatively high thermal conductivity and ease of integration. However, these materials suffer from limitations in nanoscale applications, including significant thermal losses, poor heat transport in low-dimensional systems, and mechanical instability under extreme conditions. These challenges have driven researchers to explore unconventional materials that exhibit superior thermal transport properties and are better suited to nanoscale and next-generation thermal applications.
\\

Among these unconventional options, graphene has emerged as a revolutionary material with unparalleled thermal transport capabilities. As a two-dimensional material composed of sp²-hybridized carbon atoms arranged in a honeycomb lattice, graphene combines extraordinary thermal conductivity, exceeding 3000 $Wm^{-1}K^{-1}$ at room temperature, with remarkable mechanical and electrical properties. Unlike conventional materials, graphene’s high thermal conductivity is driven by its strong covalent bonding, high phonon group velocities, and long phonon mean free paths. These unique attributes make graphene a promising candidate for heat management and energy conversion in nanoscale thermal devices.
\\

This paper presents a comprehensive study of the thermal transport properties of graphene using first-principles ab initio methods. The study provides a detailed analysis of phonon-mediated thermal conductivity in graphene, evaluates its potential as a high-performance material for thermal devices elucidating the role of phonon dispersion, scattering mechanisms, and anharmonicity in determining heat transport. 
\\

Here the concept of "velocity-lifetime trade-off" is introduced as a metric for qualitative treatment of transport properties. It also compares graphene’s thermal properties with those of conventional 2D materials to highlight its advantages and discuss potential strategies for optimizing its performance through strain engineering. The insights gained from this work aim to provide a foundation for the design and optimization of graphene-based thermal devices, addressing critical challenges in heat management and thermoelectric energy applications.

\section{Simulation Methods}
First-principles calculations for density functional theory (DFT) was conducted using projector-augmented-wave (PAW) pseudopotentials and the Perdew–Burke–Ernzerhof (PBE) exchange-correlation functional, as implemented in the Quantum Espresso code \cite{QE-2009}. To construct the unitcell, a vacuum slab with a thickness of 20 Å was added along the z-direction to minimize interlayer interactions. The plane-wave cutoff energy and wavefunction cutoff was set to 360 Ry and 45 Ry respectively, and a Monkhorst–Pack k-point mesh of $2 \times 2 \times 1$ was used. Ionic and electronic degrees of freedom were fully relaxed until the forces on all atoms were less than $10^{-5} \, \text{eV/Å}$, with the self-consistent loop convergence criterion set to $10^{-6} \, \text{eV}$.
\\

Phonon transport properties were calculated using the phonon Boltzmann transport equation (pBTE), as implemented in the Phono3py code \cite{phonopy-phono3py-JPSJ}. The calculation required harmonic and anharmonic interatomic force constants (IFCs), both of which were derived using first-principles methods alongside a cold smearing approach to ensure energy conservation.
\\


\begin{figure}[ht!]%
    \centering
    \subfloat[\centering Top view ]{{\includegraphics[width=3cm]{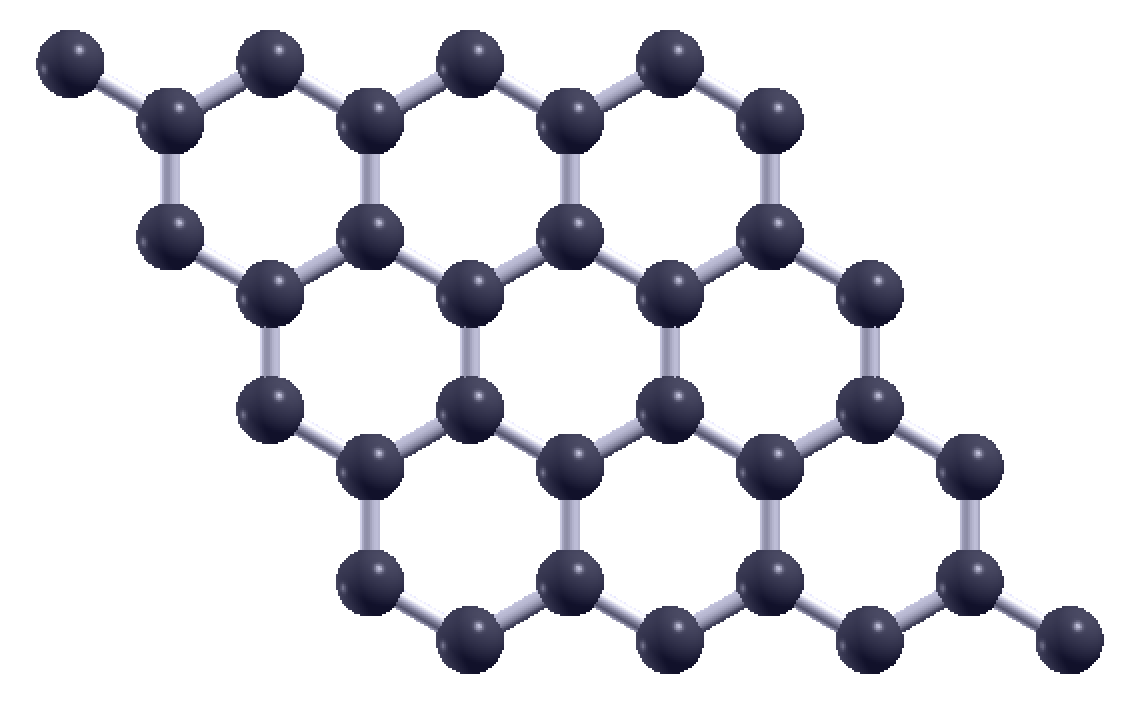} }}%
    \qquad
    \subfloat[\centering Side view]{{\includegraphics[width=3cm]{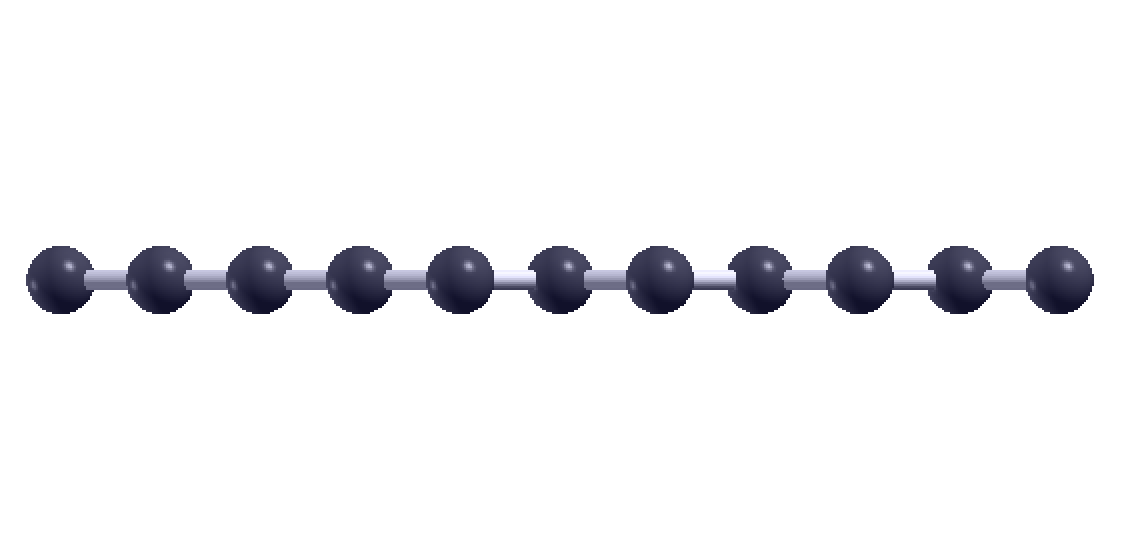} }}%
    \caption{Molecular stucture of Graphene}%
    \label{fig:structure}%
\end{figure}

\begin{figure}[h]
  \includegraphics[width= .475\textwidth]{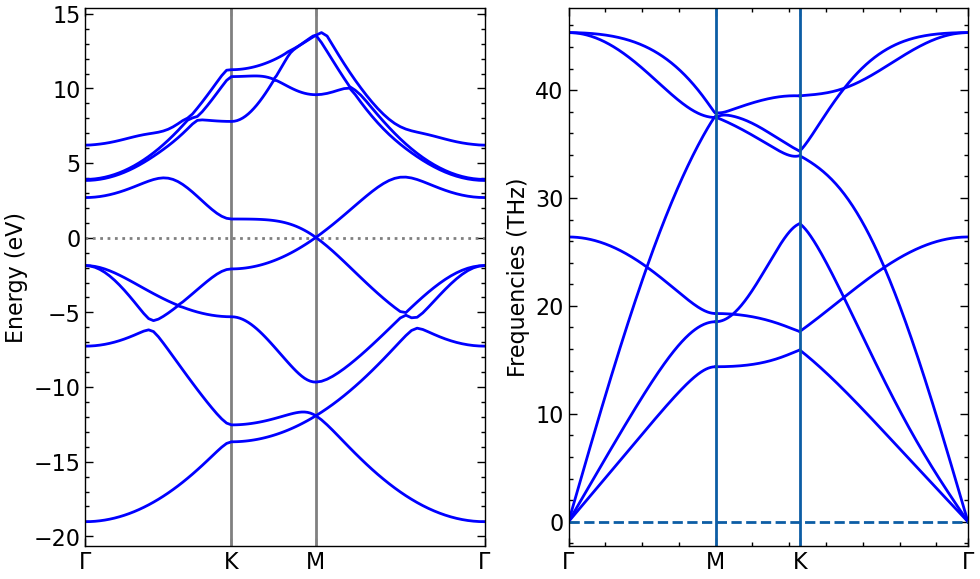}
  \label{fig:modelmm1}
  \caption{(left panel): Electronic dispersion of Graphene, (right panel): Phonon dispersion of Graphene calculateed using displacement method  }
\end{figure}

A well-converged $2 \times 2 \times 1$ supercell was adopted, and a cutoff distance of 5 Å was used to account for atomic interactions. The four-phonon scattering term is not included in these calculations, as it primarily influences high-temperature behavior or materials with inherently high lattice thermal conductivity, as supported by prior findings.
\section{Results and Discussions}
Graphene exhibits remarkable thermal transport properties, driven by its unique phonon dynamics and robust bonding. The calculated lattice constant of graphene, 2.516 \r{A}, aligns with experimental results, highlighting the strength of its covalent carbon-carbon bonds. Phonon dispersion calculations reveal a broad spectrum extending up to 45 THz, a direct result of the stiffness of the C-C bonds.

\subsection{Boltzman transport}

The lattice thermal conductivity (\(\kappa\)) of graphene is analysed, by drawing relation to it's phonon frequencies, group velocities, and scattering times. The strong covalent carbon-carbon bonds and planar symmetry result in high phonon group velocities and minimal scattering, which collectively contribute to its thermal conductivity (\(\kappa\)) exceeding 3000 $Wm^{-1}K^{-1}$ at room temperature. This positions graphene as a benchmark material for heat dissipation and energy transport in nanoscale devices.
\\

The lattice thermal conductivity (\(\kappa\)) of graphene can be rigorously described using the Boltzmann transport equation (BTE), which accounts for the contributions of all phonon modes and their interactions. The thermal conductivity is derived from the phonon-mediated heat flux and expressed as:
\begin{equation}
\kappa = \frac{\hbar^2}{\Omega k_B T^2} \sum_\nu \bar{n}_\nu (\bar{n}_\nu + 1) C_\nu \omega_\nu F_\nu,
\end{equation}
where \(\Omega\) is the unit cell volume, \(k_B\) is the Boltzmann constant, \(T\) is the temperature, and \(\bar{n}_\nu\) represents the Bose-Einstein phonon occupation number. Here, \(C_\nu\), \(\omega_\nu\), and \(F_\nu\) denote the phonon specific heat, frequency, and mean free path, respectively. The mode-resolved thermal conductivity can further be expressed as a summation over phonon modes:
\begin{equation}
\kappa = \sum_\lambda \kappa_\lambda,
\end{equation}
with each mode-specific contribution given by:
\begin{equation}
\kappa_\lambda = \tau_\lambda C_{ph}(\omega_\lambda) v_\lambda^2 \cos^2 \theta_\lambda.
\end{equation}

\begin{figure}[h!]
\centering
  \includegraphics[width= 0.475\textwidth]{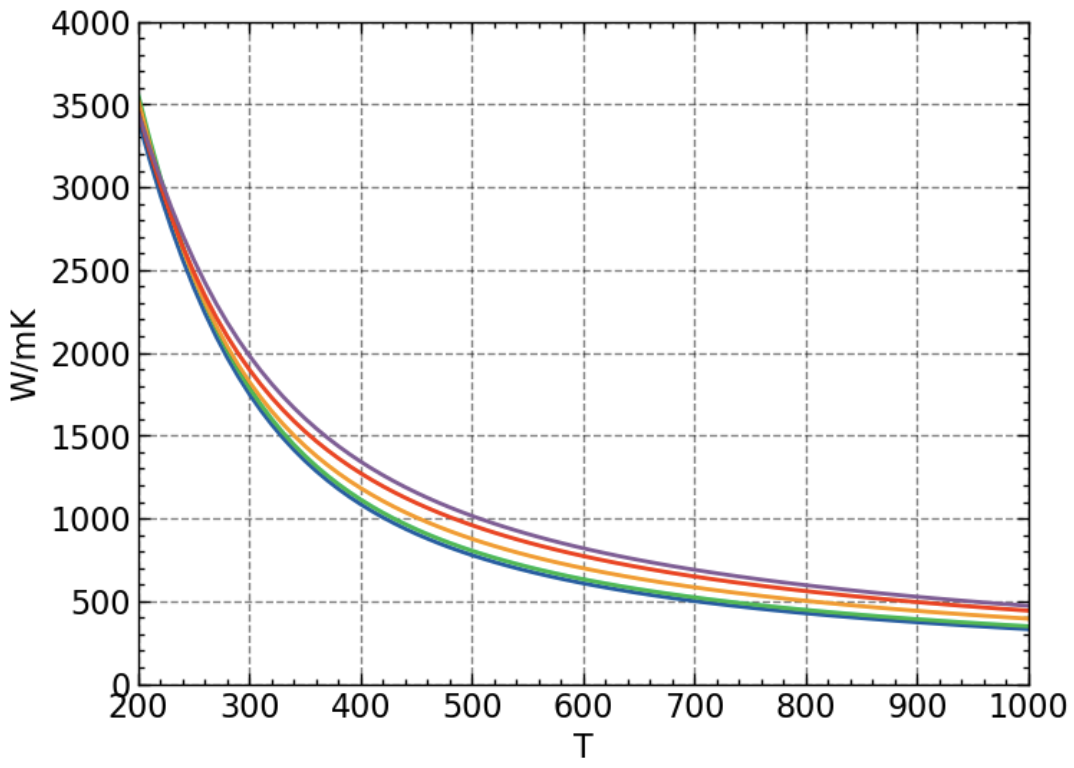}
  \caption{Lattice thermal conductivity against the temperature. The figure shows a slight deviation in the conductivity of graphene with strain: purple ( $\epsilon$ = 0),  green ( $\epsilon$ = 0.02), blue ( $\epsilon$ = 0.04), red ( $\epsilon$ = -0.02) and yellow ( $\epsilon$ = -0.04)}
  \label{fig:kappa}
\end{figure}

In this formulation, \(\tau_\lambda\) represents the phonon lifetime, \(C_{ph}(\omega_\lambda)\) is the mode-specific phonon heat capacity, and \(v_\lambda\) is the phonon group velocity. The angular term \(\cos^2 \theta_\lambda\) accounts for the directional dependence of phonon transport. The heat capacity of each phonon mode is given by:
\begin{equation}
C_{ph}(\omega) = \frac{\hbar^2 \omega^2}{k_B T^2} \frac{\exp(\hbar \omega / k_B T)}{\left[\exp(\hbar \omega / k_B T) - 1\right]^2}.
\end{equation}
These equations highlight the critical contributions of acoustic phonon modes (longitudinal acoustic, transverse acoustic, and flexural modes) to graphene's thermal conductivity.
\\

The anharmonic phonon-phonon interactions, captured by the third-order force constants (3rd FCs), play a crucial role in determining phonon lifetimes (\(\tau_\lambda\)) and thus the thermal transport properties of graphene. These interactions describe three-phonon scattering processes, where energy and momentum are conserved through phonon absorption and emission. The phonon linewidth \(\Gamma_\lambda\), which is inversely proportional to the phonon lifetime, encapsulates the scattering processes and is given by:
\begin{equation}
\Gamma_\lambda = \frac{1}{\tau_\lambda}.
\end{equation}

From \ref{fig:kappa}, the deviation in lattice thermal conductivity of graphene is much lower than other 2D materials. A strain variance of almost 0.08 results in atmost 200K deviation in the conductivity at room temperature, which is about 10\% of the total conductivity. This invariance to strain can be leveraged in many high thermal transport applications where resistance to compressive and tensile strains play a crucial role in preserving device accuracy and reliability.

\subsection{Acoustic Modes}
The thermal conductivity of Graphene is orders of magnitude higher than that of most other 2D materials. This exceptional performance is attributed to the high group velocities of acoustic phonons and the long phonon mean free path (MFP), which can exceed several micrometers. The in-plane acoustic modes, particularly the LA and TA branches , dominate heat conduction due to their high velocities and low scattering rates \ref{fig:modes}. While the ZA mode typically exhibits greater anharmonicity, its contribution to thermal conductivity remains significant due to graphene's planar symmetry, which reduces the impact of out-of-plane distortions.  
\\

The optical modes do not contribute to thermal conductivity due to their flat-band topology. As phonon group velocities are given by the derivatives of the frequencies w.r.t to their wave vectors, steeper bands (usually the acoustic modes) near the $\Gamma$ points are the primary drivers of thermal conductivity and transport, while the optical modes give very low group velocities leading to less contribution.

\subsection{Frequencies, Group velocities and Lifetimes}
The high thermal conductivity of graphene arises from the suppression of scattering and the high stiffness of its bonds, which ensures efficient heat transport.
Compared to other 2D materials, graphene consistently outperforms in thermal transport. For instance, silicene, with its increased anharmonicity and lower phonon frequencies, exhibits a thermal conductivity of less than $\kappa =$10 $Wm^{-1}K^{-1}$, while MoS$_2$, a widely studied transition metal dichalcogenide, achieves $\kappa \approx$ 83 $ Wm^{-1}K^{-1}$. These values highlight graphene’s unique ability to combine strong covalent bonding, high phonon group velocities, and minimal scattering.
\\

The third order force constants were incorporated into the BTE framework allowing for the accurate modeling of phonon lifetimes and their influence on thermal conductivities. Only the 3-phonon scattering mechanisms were considered in the calculations, as at higher temperatures, phonon-phonon scattering intensifies and 4-phonon mechanisms take over leading to further drops in lattice conductivity.
\\

\begin{figure*}[t!]
    \centering

    \begin{subfigure}[t]{.8\textwidth}
        \centering
        \includegraphics[width=\textwidth]{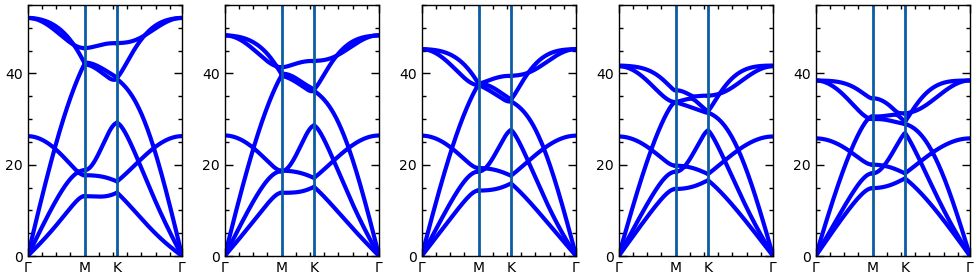}
        \caption{Phonon dispersions of mono layer graphene}
        \label{fig:freq}
        
    \end{subfigure}

    \begin{subfigure}[t]{.8\textwidth}
        \centering
        \includegraphics[width=\textwidth]{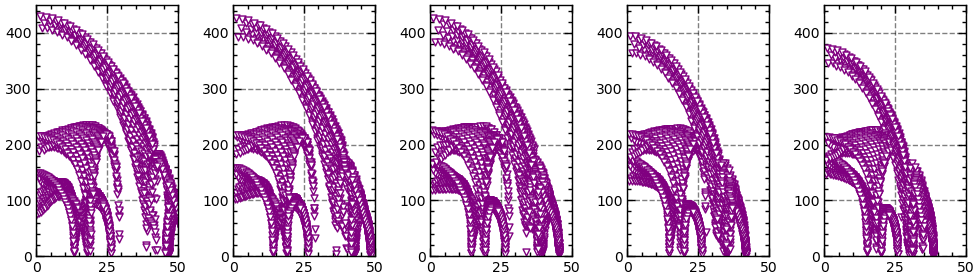}
        \caption{Group velocities}
        \label{fig:gv}
        
    \end{subfigure} 
    
    \begin{subfigure}[t]{.8\textwidth}
        \centering
        \includegraphics[width=\textwidth]{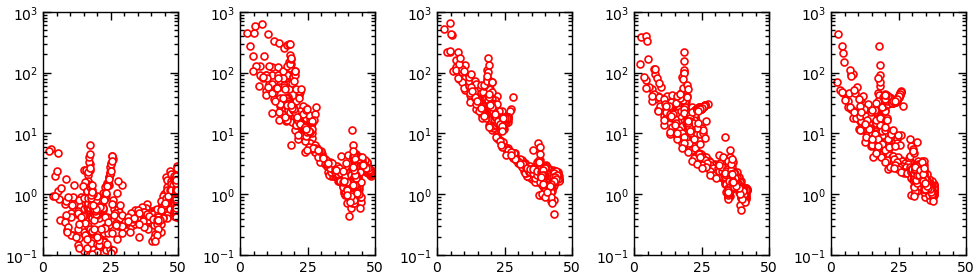}
        \caption{Phonon lifetimes}
        \label{fig:lifetimes}
        
    \end{subfigure}
    
    \begin{subfigure}[t]{.8\textwidth}
        \centering
        \includegraphics[width=\textwidth]{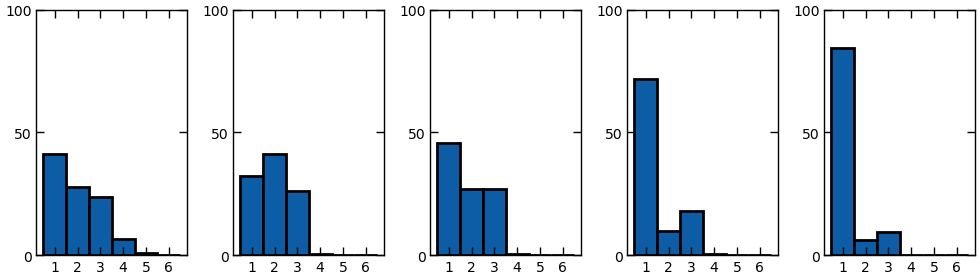}
        \caption{Contribution of each mode to total thermal conductivity. Modes numbered 1 through 6 represents the ZA, TA, LA, ZO, TO and LO modes respectively.}
        \label{fig:modes}
    \end{subfigure}
    
    \caption{Phonon properties for each value of strain (-0.04, -0.02, 0, 0.02, 0.04) shown from left to right.}
    \label{fig:phonon}
\end{figure*}

The phonon dispersion of Graphene shows a linear scaling with strain i.e. the frequencies go down with increasing tensile strain. The drop in frequencies is mostly observed in the optical modes only, while the acoustic modes stay relatively the same. However, the gap between the ZA and TA modes diminish to zero at higher values of lattice constants ($\r{A}$). Also, at high positive strains, the optical modes and even the LA and TA modes obtain a flatter topology leading to minimal contribution to thermal conductivity (as can be seen from \ref{fig:modes}) due to very low group velocities. In contrast to lower strains (compressive), as the phonon dispersions spans a larger spectrum (upto 52 THz), the bands are forced to be relatively steep leading to high V$_g$.
\\

However, the transport characteristics are optimal for the relaxed structure ( $\epsilon=0$) due to the "group velocity-lifetime trade-off". At high values of compressive strain, even though the group velocities are high, the phonon lifetimes are reduced due to greater phonon-phonon scattering \ref{fig:lifetimes}, which overall affects transport adversely.
While on the other hand at higher tensile strain, the scattering is low but due to flat bands causing lowering of V$_g$, a similar effect on thermal conductivity is observed.
\\
The above phenomena can be validated from observing the contributions of each mode towards the overall lattice thermal conductivity ($\kappa$). At negative strains, all the acoustic modes contribute equally, but upon increasing $\epsilon$, a preference for the ZA mode becomes evident.

\section{Conclusion}

While pristine graphene's gapless electronic structure limits its thermoelectric efficiency, its high thermal conductivity can be optimized for thermoelectric applications through various strategies. These include chemical doping, which alters the phonon relaxation time and modifies the density of states near the Fermi level, and strain engineering, which introduces pseudo-magnetic fields and changes the electronic and phononic transport characteristics. Heterostructures, formed by stacking graphene with other 2D materials, offer additional pathways to open a band gap and enhance thermoelectric performance. These modifications can balance the thermal and electronic transport properties, enabling graphene to be used in energy conversion and heat dissipation technologies.

The combination of solving the Boltzmann transport equation along with third-order force constants and phonon mode analysis presented in this study provides a comprehensive understanding of graphene's thermal transport mechanisms. The insights from these calculations highlight graphene's potential for next-generation thermal devices, where efficient heat dissipation is critical. 

Graphene's exceptional thermal conductivity, driven by its unique phonon properties and low anharmonic scattering, establishes it as a model material for nanoscale heat management. This study provides a foundation for leveraging graphene's capabilities in practical thermal devices and energy applications, offering pathways for further optimization and integration.



\begin{acknowledgments}
We express our sincere gratitude to the IDAPT Supercomputing Facility at the Indian Institute of Technology (IIT) BHU for providing their high-performance computing resources essential for this research, Quantum ESPRESSO for first-principles calculations, and Phonopy and Phono3py for phonon and thermal transport calculations.

\end{acknowledgments}

\nocite{*}

\bibliography{apssamp}

\end{document}